# Tunable inverse spin Hall effect in nanometer-thick platinum films by ionic gating

Sergey Dushenko[1,*,†], Masaya Hokazono[1,*], Kohji Nakamura[2], Yuichiro Ando[1], Teruya Shinjo[1], Masashi Shiraishi[1†]

[1]Department of Electronic Science and Engineering, Kyoto University, Kyoto 615-8510, Japan
[2]Department of Physics Engineering, Mie University, Mie 514-8507, Japan

[*]These authors contributed equally to the study
[†]Corresponding authors
[†]Correspondence to dushenko89@gmail.com, mshiraishi@kuee.kyoto-u.ac.jp

Electric gating can strongly modulate a wide variety of physical properties in semiconductors and insulators, such as significant changes of conductivity in silicon, appearance of superconductivity in $SrTiO_3$, the paramagnet-ferromagnet transition in (In,Mn)As and so on. The key to such modulation is charge accumulation in solids. Thus, it has been believed that such modulation is out of reach for conventional metals where the number of carriers is too large. However, success in tuning the Curie temperature of ultrathin cobalt gave hope of finally achieving such degree of control even in metallic materials. Here, we show reversible modulation of up to two orders of magnitude of the inverse spin Hall effect—a phenomenon that governs interconversion between spin and charge currents—in ultrathin platinum. Spin-to-charge conversion enables the generation and use of electric and spin currents in the same device, which is crucial for the future of spintronics and electronics.

## Introduction

Electric gating—as an instrument to modulate properties of materials via control of carrier density—became famous in the middle of 20th century, when research of W. Shockley, W.H. Brattain and J. Bardeen was lauded by the Nobel prize in Physics in 1956. Nowadays, it lies in the technological foundation of our civilization—field effect transistor, where electric gating controls semiconductor channel and switches device between on- and off-states. However, application of electric gating techniques for a long time was restricted to semiconducting materials, where carrier density is low enough to be tuned by the gate voltage. Nowadays, electric gating is also used to control devices in the emerging field of two-dimensional materials like graphene and transition metal dichalcogenide monolayers. While carrier density for two-dimensional materials can be high, carriers are located at the interface—region where electric gating has the most influence. In contrast to the cases described above, metals are bulk materials, and have high carrier density at the same time. Thus, significant modulation of properties of metals by electric gating was mostly considered out of reach. While electric gating controls carrier density and directly influences conductivity of the material, it can be also applied to control any property of the material that depends on the position of the Fermi level. A few prominent examples include electric gating controlled insulator-superconductor[1] and paramagnet-ferromagnet transitions[2]. Lack of such degree of control over physical phenomena in metals, presented a big disadvantage to functionality of metallic devices. However, success in the tuning of the Curie temperature of ultrathin Co gave hope of finally achieving such degree of control even in metallic materials.[3]

In 2006 Saitoh et al. converted the pure spin current into the electric charge current using the inverse spin Hall effect (ISHE) in metallic Pt layer,[4] while Valenzuela et al. detected the same effect in the aluminum channel of lateral spin valve.[5] The ISHE originates from the spin-orbit interaction inside the material. Due to the spin-orbit interaction the scattering direction of the carriers depends on their spin direction. Thus, the ISHE couples spin current with the electric charge current and allows their interconversion: the longitudinal pure spin current generates transversal charge current (the ISHE) and vice versa (the spin Hall effect). After the initial prediction[6] and experimental detection[7] it took almost 20 years until the importance of the effect was recognized by scientific community. Pt became the dominant material choice to introduce the spin-charge conversion in studied systems due to its large spin Hall angle (which characterizes the efficiency of the conversion between spin and charge currents) and easy fabrication.[8–11] In recent years, novel spin-charge conversion effects like spin-charge conversion due to the electric field and spin-orbit coupling at the interface between two materials,[12] or spin-charge conversion via the spin-momentum locking in topological insulators[13–15] challenged the dominance of the ISHE generated by Pt and other heavy metals.[16] However, though some of the aforementioned systems possess higher spin-charge conversion efficiency than ISHE in Pt, at the moment they are very sensitive to interface quality or not robust at room temperature. Through the years, there were many attempts to achieve control over the spin-charge conversion process. Such control was achieved using electric bias tuning of Schottky barrier in semiconductors[17] and electric gating in the various two-dimensional systems[18–21]. However, as discussed above, in contrast to semiconductors and two-dimensional systems, electric gating control over spin-charge conversion via ISHE in metals remained a formidable challenge. Recent studies showed that it is possible to change the ISHE of Pt through composition control of the sample: by either adjusting number of the scattering centers[22,23], or substituting part of Pt atoms with another element.[24,25] However, tuning of the ISHE in the heavy metals within a single device remained elusive so far.

In this paper, we report the largest to date modulation of the metal resistivity in ultrathin Pt film through the careful control of Pt thickness and an ionic gate technique. We show that such control over the carrier density allowed us to tune reversibly and reproducibly the amplitude of the ISHE in Pt over two orders of magnitude—a result that can be used in spin-torque and other spintronics devices that use spin-charge conversion.

## Results

### Resistivity and carrier modulation measurements

Figure 1 shows the dependence of the Pt resistivity $\rho_{Pt}$ in logarithmic scale on the inverse thickness $1/d_{Pt}$, and thickness $d_{Pt}$ (**a** and **b**, respectively), where each filled circle represents an individual sample. The solid blue line shows the resistivity calculated using Eq. 23 from the literature[26] (which takes into account scattering by both grain boundaries and film surfaces) with values $p = 0.8$—fraction of carriers specularly scattered at the surface of Pt layer, bulk resistivity $\rho_\infty = 40$ μΩ cm, mean free path $\lambda_{mfp} = 10$ nm and the grain boundary penetration parameter $\zeta = 0.25$.[26] Experimental data follows theoretical calculation, which shows that Pt samples were successfully fabricated down to the smallest thickness. Sharp increase in the Pt resistivity with the decreasing thickness is due to the increased contribution from the surface and grain boundaries scattering in the thin layers. The atomic force microscopy measurements (Supplementary Note 11) also confirmed continuous nature of the Pt films.

Figure 2 shows the schematic image of the samples used in the study. Thin Pt films (in the thickness range from 1.5 nm to 20 nm) were grown on top of the insulating Gadolinium Gallium Garnet (GGG)/Yttrium Iron Garnet (YIG) substrates. We modulated charge carrier density in our Pt devices with electric top gate using the ionic liquid technique, which is also commonly referred to as the electric double-layer transistor method. During the measurement, a sample with the gate was mounted in vertical position into the cavity of the electron spin resonance system. Figure 3 shows the ratio of the resistivity of the samples measured at gate voltage $V_G = -2V$ and $V_G = +2V$: $k = R(V_G = -2V)/R(V_G = +2V)$. Simple calculation using the free electron Drude theory of metals that assumes one free electron per atom gives estimation of the atom and carrier density in Pt $n = 6.6 \cdot 10^{22}$ cm$^{-3}$. However, band calculations predict smaller number of 0.4 6s-band electrons per Pt atom.[27] Experimentally even lower values of 0.24 conduction electrons per Pt atom were measured in thin Pt films, with bulk carrier density calculated to be $n = 1.6 \cdot 10^{22}$ cm$^{-3}$.[28] While Pt is a two-band conductor with dominating carriers from the closed s-like Γ-electron and open d-like X-hole Fermi surfaces,[27,29,30] the difference in effective mass and specular reflection of light electron carriers and heavy hole carriers can lead to electron band dominated conduction in thin films.[28,31] Using formula $n = \sqrt{\frac{3}{8\pi}} \left( \frac{\sigma_\infty}{\lambda_{mfp}} \frac{h}{e^2} \right)^{3/2}$, where e—elementary charge, h—Planck constant, and values of $\sigma_\infty$ and $\lambda_{mfp}$ obtained from the thickness dependence of resistivity, value of the bulk carrier density in our Pt is estimated to be $n = 6·10^{21}$ cm$^{-3}$, which is the same with result calculated for thin Pt films from the data reported in the literature.[26] Carrier density—modulated by the gate voltage in 2 nm-thick sample—is estimated to be $n = 4.2·10^{21}$ cm$^{-3}$ at $V_G$ = -2 V, and $n = 7.9·10^{21}$ cm$^{-3}$ at $V_G$ = 2 V. Thus, induced by the ionic gel carrier density is $\Delta n = ±2·10^{21}$ cm$^{-3}$, which gives induced sheet carrier density of $\Delta n_{sh} = \Delta n · d_{Pt} = 4·10^{14}$ cm$^{-2}$. This value is within the range of commonly reported carrier density modulation using the ionic liquid: the order of $10^{14}$ carriers/cm$^2$ is routinely achieved, with the highest reported values larger than $10^{15}$ carriers/cm$^2$.[32] Using relation between $n$ and $\sigma$ from the above, we can theoretically estimate $k = \frac{R|_{V_G = -2V}}{R|_{V_G = +2V}} = \left(\frac{d_{Pt}+\Delta n_{sh}/n}{d_{Pt}-\Delta n_{sh}/n}\right)^{2/3}$, where $n = 6·10^{21}$ cm$^{-3}$. Figure 3 shows experimentally measured $k$ for samples with various Pt thickness (purple filled circles), and theoretically estimated $k$ assuming the same induced sheet carrier density in all devices $\Delta n_{sh} = 4.8·10^{14}$ cm$^{-2}$ (blue line). In agreement with theoretical calculation, all devices showed increased resistivity modulation factor $k$ with the decreased thickness (see Supplementary Notes 4, 5, 9, 14 for further discussion on the resistivity modulation using ionic gel, and Supplementary Note 15 for detailed discussion on the effect of the charge screening on gate modulation). Additionally, in thin films resistivity contribution due to the grain hopping can possibly be modulated by the gate voltage application. However, measurements of the temperature dependence of the resistivity indicate that its contribution at 250 K (temperature of the spin pumping and ISHE measurements) is on the order of only a few percent even in 2 nm-thick films (see Supplementary Notes 16 for detailed discussion). We achieved the resistivity modulation $k$ in our samples up to 280%: this is more than one order of magnitude larger than previously reported in other studies.[31,33–35]

**Spin-charge conversion measurements**

For the ISHE measurements, after setting the gate voltage, we cooled the sample from room temperature to 250 K. Ionic leak current, comparable to the spin-charge conversion current at room temperature in Pt (on the order of nA), is completely suppressed at 250 K, when ion molecules in the ionic gel become immobile. For the details about ionic gel preparation and measurement procedure, see Methods section. Magnetic field of the microwaves applied to the cavity with sample drives magnetization of the ferrimagnet YIG layer into precession at the certain value of the external magnetic field, known as the ferromagnetic resonance field $\mathbf{H}_{FMR}$.[36] Precession of the magnetization induces transfer of the angular momentum from YIG into the adjacent Pt layer without any charge transfer, i.e. generates pure out-of-plane spin current $\mathbf{j}_s$, where spin direction is determined by the direction of the applied static magnetic field. This pure spin current generation method is commonly referred to as the spin pumping.[37,38] The pure out-of-plane spin current is converted into the in-plane charge current via the ISHE, which is measured at Ti/Au electrodes located at the opposite sides of the samples. Figure 4d shows the electromotive force generated in the 2 nm-thick Pt sample under the gate voltage of 0 V during the microwave absorption in YIG layer. The generated pure spin and charge currents are proportional to the absorbed microwave power,[39] thus generated electromotive force follows the Lorentzian shape of the microwave absorption spectrum. Direction of the injected spins $\mathbf{\sigma}$ is reversed together with the direction of the external magnetic field (characterized by angle $\theta_H$), which results in the sign change of the spin-charge conversion current: $\mathbf{j}_c \propto \mathbf{j}_s \times \boldsymbol{\sigma}$. In agreement with the ISHE theory, sign of the generated electromotive force was reversed between $\theta_H = 0°$ (blue filled circles) and $\theta_H = 180°$ (purple filled circles). Subtracting electromotive force data for the opposite directions of the external magnetic field removes spurious contributions independent of magnetic field (for example, the Seebeck effect). The amplitude of the ISHE voltage was extracted from the fitting of the magnetic field-averaged electromotive force ($V_{\theta H=0°}$ - $V_{\theta H=180°}$)/2 using symmetrical and asymmetrical Lorentzian components (see Supplementary Notes 6-8, 13 for more details, examples of the averaging and fitting procedure). Figures 4a and 4b show change in the amplitude of the current and voltage generated by the ISHE in the 2 nm-thick Pt sample. In contrast to the microwave absorption spectrum (Fig. 4i-4n), electromotive force generated via the ISHE (Fig. 4c-4h) was strongly modulated with the application of the gate voltage. The amplitude of the spin-charge conversion current was tuned from the $I_{ISHE} = V_{ISHE}/R = 3.0$ nA at $V_G = -2$ V to $I_{ISHE} = 0.1$ nA at $V_G = 2$ V. In our best sample, we achieve modulation of the $V_{ISHE}$ and $I_{ISHE}$ from 100% at $V_G = -2$ V down to 0.8% and 1.7%, respectively, at $V_G = 2$ V. Figure 5c shows reproducibility of the modulation of the spin-charge conversion current in two different sweeps of the gate voltage for the same sample and in the different sample with the same Pt thickness $d_{Pt} = 2$ nm. These results show the successful achievement of control over the ISHE in a metallic material.

**Discussion**

In the following paragraph, we show that the ISHE observed in our samples is intrinsic in nature and originates from the inter-d-band excitations. The ISHE spin-charge conversion current is given by the equation[40], $I_{ISHE} = w\theta_{SH}\lambda_s\tanh(d/(2\lambda_s))(2e/\hbar)j_s^0$, where $w$—channel width (same for all samples), $\theta_{SH}$—spin Hall angle, $\lambda_s$—spin diffusion length, $d$—channel thickness, $\hbar$—reduced Planck constant, $j_s^0$—injected spin current density at the YIG/Pt interface. Also, due to the Elliott-Yafet spin relaxation mechanism[41,42] in Pt $\sigma_{SH} = \sigma\theta_{SH} \propto \lambda_s\theta_{SH}$, where $\sigma_{SH}$ and $\sigma$ are spin Hall and electrical conductivities, respectively.[22,23,43] Disregarding small changes in the factor $\tanh(d/2\lambda_s)$ between different samples (it is close to 1 for all channel thicknesses because of the scaling of the spin diffusion length with resistivity of Pt samples[22,23,43]), one arrives at: $I_{ISHE} \cong A\sigma_{SH}j_s^0$, where $A = w(\lambda_s/\sigma)\tanh(d/(2\lambda_s))(2e/\hbar)$ can be considered as a constant across the samples. Since YIG surface treatment and sputtering conditions for Pt were identical across all samples, we can assume similar injected spin current density $j_s^0$ at the Pt/YIG interface. Thus, generated spin-charge conversion current $I_{ISHE}$ from sample to sample was solely controlled by the spin Hall conductivity (SHC) $\sigma_{SH}$ of the sample: $I_{ISHE} \cong A`\sigma_{SH}$, where $A` = Aj_s^0$ is a constant. The ISHE in Pt was theoretically predicted[8,44,45] and experimentally confirmed[10,11] to be dominated by the intrinsic mechanism, unless the superclean regime is entered ($\rho_{Pt} < 15$ μΩ cm), where extrinsic ISHE cannot be neglected anymore.[23] Large spin-orbit splitting lifts the double degeneracy of the d-bands near the L and X points at the Fermi level in Pt. Intrinsic ISHE originates from the interband scattering between these orbitals,[45] and—according to the band calculations—SHC should exhibit sharp decrease with increasing resistivity.[8,44] In contrast, in the previous experimental studies, SHC was measured to be independent of Pt resistivity.[22,23,46] While the spin Hall angle of Pt varies greatly among different studies, it was shown to originate from the linear scaling of the spin Hall angle with the resistivity of the sample[22,23,43], leaving SHC $\sigma_{SH} = \sigma\theta_{SH}$ unaffected by changes in the resistivity. Thus, to the best of our knowledge, the theoretically predicted dependence of the SHC on the resistivity of material has been never observed experimentally neither in Pt, nor in other materials. Our high-resistivity samples address this discrepancy between the theory and experiment, and access the resistivity-dependent regime of SHC. All experimental studies so far were carried out on the low-resistivity Pt samples, where interband excitations that govern SHC are controlled by the $\hbar/\Delta$, where $\Delta$ is the spin-orbit-induced splitting of d-bands. In contrast to the low-resistivity regime ($\rho_{Pt} < 40$ μΩ cm), in the high-resistivity regime ($\gamma \gg \Delta$) interband excitations are governed by the quasiparticle lifetime $\hbar/\gamma$, which is roughly inversely proportional to resistivity. Figure 6 shows dependence of the $\sigma_{SH}(\rho) \cong I_{ISHE}(\rho)/A$` (calculated using $I_{ISHE}$ measured at $V_G = 0$ V and $A$`$= 0.05$ nA Ω cm) on the resistivity of the samples, which was controlled by the thickness of the Pt layer (see Fig. 1). SHC showed strong decrease with the resistivity in our samples that followed the $\rho^{-2}$ dependence (Fig. 6 dashed blue line) theoretically predicted for the $\sigma_{SH}$.[8,44] Our results provide experimental evidence for the inter-d-band excitations origin of SHC in Pt.

Keeping in mind the inter-d-band transitions nature of the SHC, we discuss a possible mechanism of the observed strong suppression of the ISHE at $V_G > 0$. As discussed above, in thin films contribution of the s-like electrons to conduction is dominant, while contribution from the d-like carriers is small due to the specular reflection and large effective mass. However, the main scattering mechanism for the s-like conduction electrons is phonon-induced scattering into d-like empty states at the Fermi level, which depends strongly on the density of the d-states. At the same time, the density of states of the d-bands affects the SHC governed by the inter-d-band transitions. Interestingly, band calculations showed that density of the d-states in Pt sharply decreases above the Fermi level.[47] Thus, the upshift of the Fermi level at positive $V_G$ should lead to decrease in the resistivity due to the increased number of electrons at Γ point and the decreased scattering rate through the d-states, and decrease in the SHC due to the decreased inter-d-band scattering because of the moving away from the points where the split d-bands are close to each other and the decreased number of d-states. This is consistent with our experimentally observed results where the decrease in the sample resistance is followed by the decrease in the spin-charge conversion current generated through the ISHE (Fig. 5). Such reduction of the SHC with the tuning of the Fermi level was predicted for the bulk Pt, though large suppression of the SHC was estimated to occur on the shift of Fermi level on the order of 1 eV,[45] which is larger than expected in our case. We hope that our results will motivate theoretical studies on the Fermi level dependence of the SHC in ultrathin Pt films, where increased scattering and lower carrier density, together with the few-atomic layer thickness of the film, can lead to a difference in the SHC in comparison with bulk Pt, which can help to explain the sharper dependence of the SHC on the position of the Fermi level.

Finally, we note that the anomalous Hall effect (AHE) induced by the proximity to ferrimagnetic insulators[48] or gate voltage application was reported in thin Pt films.[49] Such induced magnetic proximity effect can lead to a reduction of the SHC in Pt.[50] While the mechanism of the induced magnetic moments that causes the AHE in Pt is still not completely understood, the AHE in Pt was only present at low temperatures below 200 K, and a large part of the gate-induced resistance and the AHE modulation was irreversible.[48,49] In contrast, we show reversible control over the resistance and the ISHE at $T = 250$ K (see Supplementary Note 5 and Fig. 5c). Hence, magnetically induced moments in Pt are expected to emerge at temperature lower than used in our experiments. To confirm this, we carried out Hall measurements in our samples. We observed clear non-linear component that can be attributed to the AHE only at 10 K (Supplementary Note 12). Moreover, the negative magnetoresistance in Pt was also attributed to the emergence of the magnetic moments.[49] We observe switching from the positive to the negative magnetoresistance in $d_{Pt} = 2$ nm sample only at 10 K. Thus, magnetic effects appear in our system at much lower temperature than the 250 K, at which spin pumping and spin-charge conversion experiments were performed. In other study, change in sign of the spin-charge conversion in Pt-based spin-torque structures was reported with the thickness of the Pt layer.[51,52] However, it originated from the spin-charge conversion at the interface between Pt and oxidized CoFeB layer, which is absent in our case.

Our results provide insight in the fascinating physics of ultrathin Pt films and spin-charge conversion. Through the ionic gel gate, we demonstrated reversible control over the resistivity of Pt film that is one order of magnitude larger than was achieved in previous studies. Such control over the carrier density allowed us to tune the ISHE in Pt by two orders of magnitude—a result that can be used in the gate-tunable spin-charge converters, spin-torque and other types of spintronics devices. For example, it opens an exciting possibility of the gate-controlled spin-orbit torque magnetoresistive random-access memory (SOT-MRAM), where the spin current generated by the spin-charge conversion in heavy metal exerts a torque on the free magnetic layer.[53]

**Methods**

**Sample fabrication procedure**

Below is the description of the preparation and measurement procedure for each YIG/Pt sample used in the study. The GGG/YIG (1.3 μm-thick, 3 mm-long and 1 mm-wide) (Granopt, Japan) substrate was polished with agglomerate-free alumina polishing suspension (50 nm particle size), and then annealed at 1000°C in the air atmosphere for 90 min. The Pt layer was sputtered on top of YIG in Ar plasma at a rate 0.6 Å/s. Afterwards, the Ti(5 nm)/Au(100 nm) electric pads were formed on the sides of the sample by the electron beam evaporation. Ionic gel was prepared using mixture with weight ratio 9.3:0.7:20 of the PS-PMMA-PS polymer (Polymer Source, USA), DEME-TFSI ionic liquid (Kanto Chemical, Japan) and Ethyl Propionate ($CH_3CH_2COOC_2H_5$, Nacalai Tesque, Japan). Insulating double-side adhesive tape was placed on the sides of the Pt channel (inside the area covered by Ti/Au electric pads) to provide additional mechanical support for the gate electrode film, on top of which it was placed. Gate electrode film was mounted after the application of the ionic gel and was located directly above the Pt channel. See Supplementary Notes 1, 17 for further details on sample structure and fabrication.

**Measurement procedure**

For the measurement, sample was mounted in the $TE_{011}$ cavity of the electron spin resonance system (JEOL JES-FA200). The applied microwave power was set to 1 mW, and the microwave frequency to $f = 9.12$ GHz. Gate voltage was set at room temperature; after the development of the electric double layer in the ionic gel, sample was cooled to 250 K and $I$-$V$ characteristics, FMR and ISHE measurements were carried out. Constant nitrogen gas flow was supplied to the cavity with sample, which was only stopped during the refilling of the liquid nitrogen vessel. Schematic layout of the measurement procedure can be also found in Supplementary Note 2.

**Acknowledgments**

This work was supported in part by MEXT (Innovative Area "Nano Spin Conversion Science" KAKENHI No. 26103003), Scientific Research (S) "Semiconductor Spincurrentronics" (No. 16H06330) and Grant-in-Aid for Young Scientists(A) No. 16H06089. S.D. acknowledges support by JSPS Postdoctoral Fellowship and JSPS KAKENHI Grant No. 16F16064.

Authors are grateful to T. Takenobu and J. Pu for the advice on the ionic gel preparation and application.

**Authors Contributions**

S.D. and M.S. designed and supervised the experiment; M.H. prepared the samples and carried out the measurements; S.D. guided the measurements, processed and analyzed the data, and wrote the manuscript; all of the authors contributed to the discussion of the results.

**Additional information**

Supplementary Information is available online.

**Competing interests**

The authors declare no competing interests.

**Data availability**

Data measured or analyzed during this study are available from the corresponding authors on reasonable request.

**Figures**

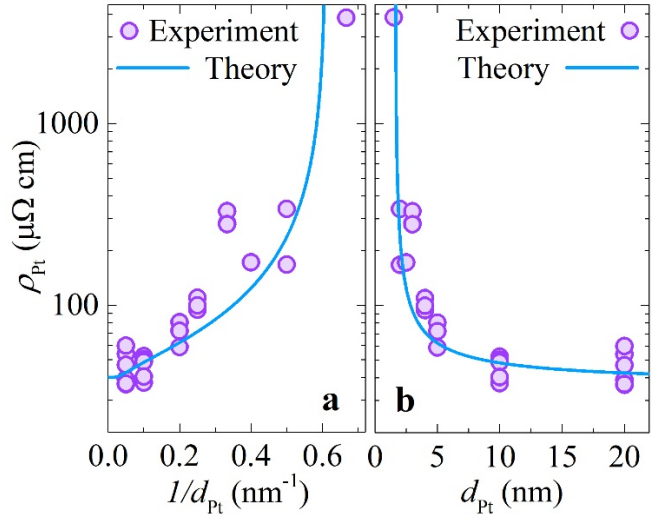

**Fig. 1 | Thickness dependence of the Pt resistivity.** Dependence of the Pt resistivity $\rho_{Pt}$ on **a**, the inverse thickness $1/d_{Pt}$, **b**, the thickness $d_{Pt}$. Note the logarithmic scale of *y*-axis. Filled circles show the experimental data, line shows the resistivity calculated using Eq. 23 from the literature[26] and values $\rho_\infty = 40$ μΩ cm, $\lambda = 10$ nm, $p = 0.8$, $\zeta = 0.25$.[26] Resistivity of the Pt films increased with the decreasing thickness due to increased contribution from the surface and grain boundaries scattering. See Supplementary Note 3 for further details.

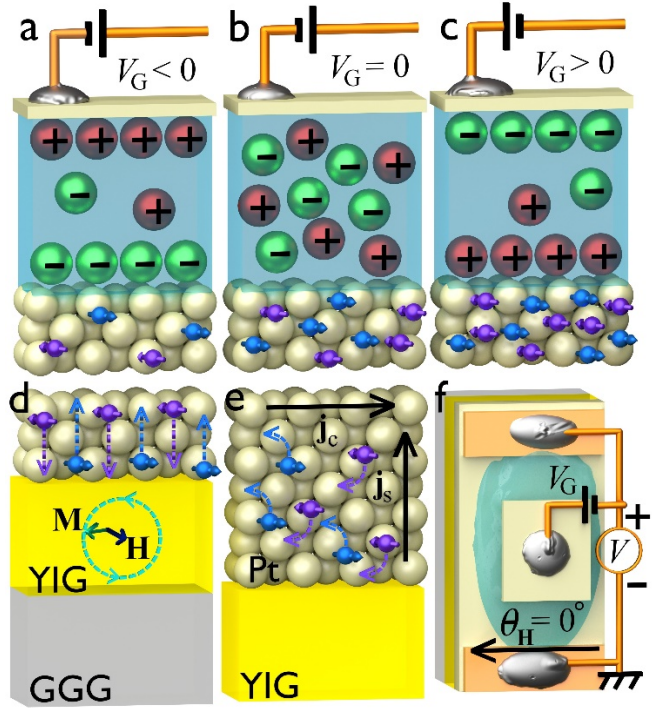

**Fig. 2 | Layout of the experiment**. **a-c**, Schematic representation of the carrier density modulation in Pt channel by using top ionic gel gate. **a**, At the negative gate voltage $V_G < 0$ ions inside the ionic gel form negatively charged layer at the interface with Pt, thus decreasing the number of electron carriers available in the channel; **b**, at $V_G = 0$ ionic gel is unordered and carrier density in the channel is not modulated; **c**, at $V_G > 0$ positively charged layer is formed at the interface with Pt leading to increased number of electron carriers in the Pt channel and decreased resistance. **d**, Under the ferromagnetic resonance condition of YIG layer, out-of-plane spin current $\mathbf{j}_s$ is injected into Pt channel via spin pumping. **e**, Inverse spin Hall effect, ISHE, inside the Pt channel converts out-of-plane spin current $\mathbf{j}_s$ into the in-plane charge current $\mathbf{j}_c$. **f**, Schematic top view of the sample. Electromotive force generated by the ISHE is detected from the Ti/Au electrodes at the ends of the sample. See Supplementary Notes 1, 17 for further details on sample structure and fabrication.

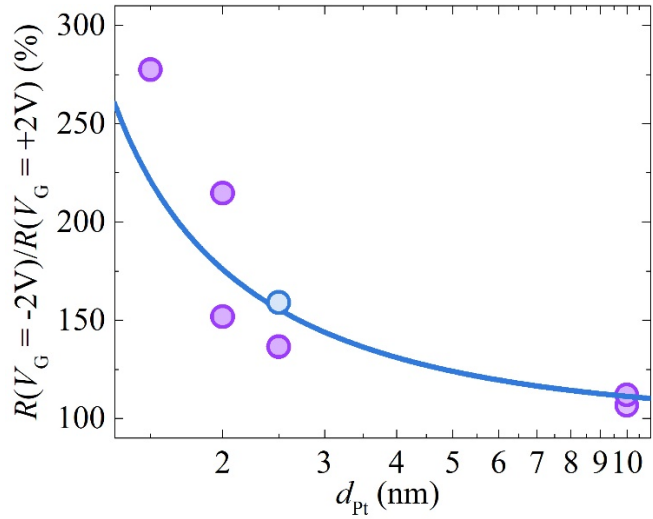

**Fig. 3 | Thickness dependence of the Pt resistivity modulation.** Dependence of the resistivity modulation factor $k = R(V_G = -2V)/R(V_G = +2V)$ on the thickness of Pt layer $d_{Pt}$. Blue solid line shows theoretical calculation assuming the same number of carriers induced in all devices, as described in the main text; purple filled circles are experimental data, each circle corresponds to a separate sample. The blue filled circle is the averaged resistivity modulation $k$ obtained for 2.5 nm sample from 15 gate voltage sweeps (see Supplementary Note 5 for details). In agreement with the theoretical calculation, devices showed increased resistivity modulation factor $k$ with the decreased thickness. Note the logarithmic scale of $x$-axis.

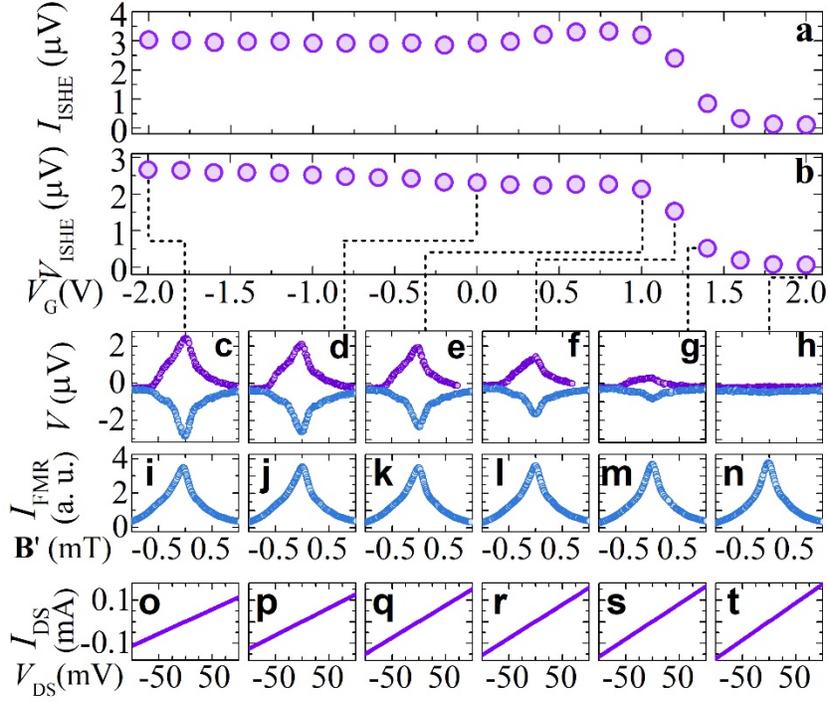

**Fig. 4 | Gate modulation data for the 2 nm-thick Pt sample. a**, Dependence of the ISHE current amplitude, $I_{ISHE}$, on the applied gate voltage $V_G$. **b**, Dependence of the ISHE voltage amplitude, $V_{ISHE}$, on the applied gate voltage $V_G$. **c-h**, Electromotive force measured for the direction of the external magnetic field $\theta_H = 0°$ (blue filled circles) and $\theta_H = 180°$ (purple filled circles), dashed black lines show corresponding experimental data points in panel **b** ($V_{ISHE}(V_G)$); **i-n**, microwave absorbance spectrum at the ferromagnetic resonance; and, **o-t**, drain-source current $I_{DS}$ dependence on the drain-source current voltage $V_{DS}$; at a set gate voltage $V_G$, where **B'**=$\mu_0$(**H**-**H**$_{Center}$). **c,i,o**, $V_G$ = -2.0 V; **d,j,p**, $V_G$ = 0 V; **e,k,q**, $V_G$ = 1.0 V; **f,l,r**, $V_G$ = 1.2 V; **g,m,s**, $V_G$ = 1.4 V; **h,n,t**, $V_G$ = 2.0 V. In contrast to the microwave absorption spectrum (**i-n**), electromotive force generated via the ISHE (**c-h**) was strongly modulated with the application of the gate voltage.

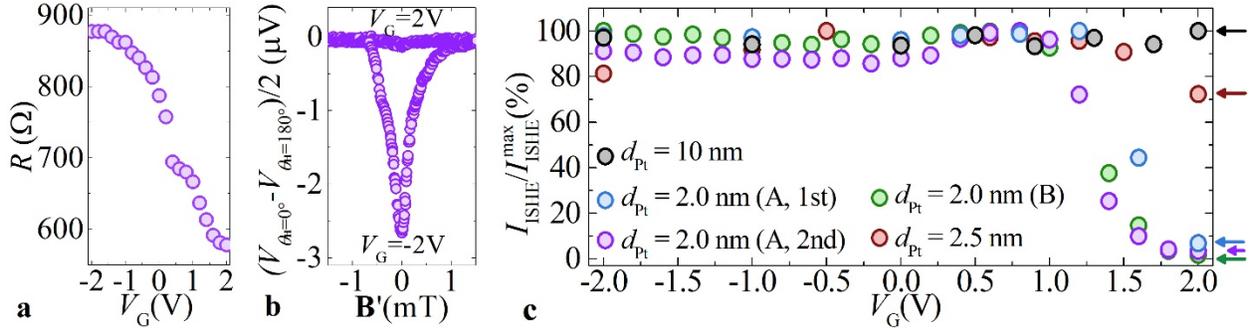

**Fig. 5 | Gate control of the spin-charge conversion in thin Pt films. a**, Resistance modulation under application of the gate voltage $V_G$ for the $d_{Pt}$ = 2 nm sample. **b**, Electromotive force detected from $d_{Pt}$ = 2 nm sample for $V_G$ = -2.0 V and $V_G$ = 2.0 V averaged over opposite directions of the external magnetic field ($\theta_H$ = 0° and $\theta_H$ = 180°) to remove spurious contributions; **B'**=$\mu_0$(**H**-**H**$_{Center}$). **c**, Comparison of the normalized spin-charge conversion current $I_{ISHE}/I_{ISHE}^{max}$ between different devices, where $I_{ISHE} = V_{ISHE}/R$ — amplitude of the generated via the ISHE spin-charge conversion current. Blue filled circles—Device A with $d_{Pt}$ = 2 nm, purple filled circles—Device A with $d_{Pt}$ = 2 nm remeasured, green filled circles—Device B with $d_{Pt}$ = 2 nm ($V_{ISHE}(V_G)$ and $R(V_G)$ can be found in Supplementary Note 10), red filled circles—device with $d_{Pt}$ = 2.5 nm, grey filled circles—device with $d_{Pt}$ = 10 nm. Comparing to the 10 nm and 2.5 nm devices, 2.0 nm devices showed larger modulation of the ISHE current, as expected from the carrier density modulation mechanism.

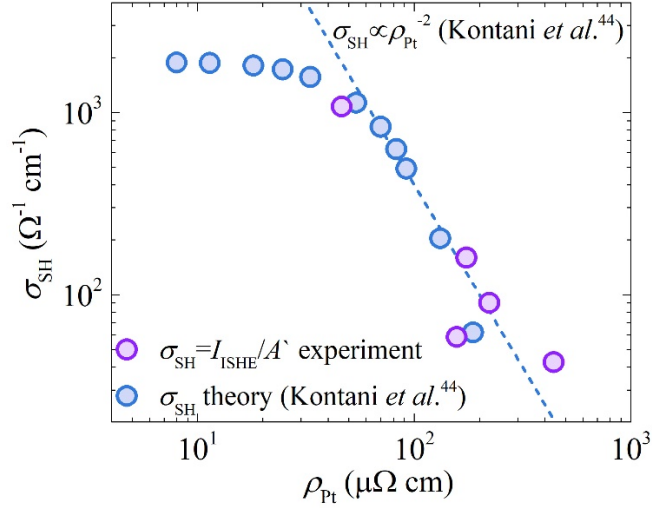

**Fig. 6 | Dependence of the spin Hall conductivity on the resistivity of Pt.** Purple filled circles show the spin Hall conductivity (SHC) calculated from the experimentally measured spin-charge conversion current $I_{ISHE}$, assuming $A` = 0.05$ nA $\Omega$ cm; blue filled circles show SHC $\sigma_{SH}$ from the band calculations of Kontani et al.[44] The experimentally measured decrease of SHC follows the theoretically predicted $\rho^{-2}$ dependence of the SHC in the high-resistivity regime (dashed blue line).[8,44] The SHC values are given in $\hbar/e$ units.